# Pulsed photoelectric coherent manipulation and detection of NV centre spins in diamond

*Michal Gulka[a,b], Emilie Bourgeois[a,c], Jaroslav Hruby[a], Petr Siyushev[d], Georg Wachter[e], Friedrich Aumayr[f], Philip R. Hemmer[g], Adam Gali[h,i], Fedor Jelezko[d], Michael Trupke[e], Milos Nesladek[a,c]*

Corresponding author: Milos Nesladek (milos.nesladek@uhasselt.be)

[a]Institute for Materials Research (IMO), Hasselt University, Wetenschapspark 1, B-3590 Diepenbeek, Belgium.
[b]Faculty of Biomedical Engineering, Czech Technical University in Prague, Sítna sq. 3105, 27201 Kladno, Czech Republic.
[c]IMOMEC division, IMEC, Wetenschapspark 1, B-3590 Diepenbeek, Belgium.
[d]Institute for Quantum Optics and IQST, Ulm University, Albert-Einstein-Allee 11, D-89081 Ulm, Germany.
[e]Vienna Center for Quantum Science and Technology, Atominstitut, TU Wien, 1020 Vienna, Austria.
[f]Institute of Applied Physics, TU Wien, Wiedner Hauptstr. 8-10, 1040 Vienna, Austria.
[g]Department of Electrical and Computer Engineering, Texas A&M University, College Station, TX 77843, USA.
[h]Institute for Solid State Physics and Optics, Wigner Research Centre for Physics, Hungarian Academy of Sciences, PO Box 49, H-1525 Budapest, Hungary.
[i]Department of Atomic Physics, Budapest University of Technology and Economics, Budafoki út 8, H-1111 Budapest, Hungary.

**Abstract:**

Hybrid photoelectric detection of NV magnetic resonances (PDMR) is anticipated to lead to scalable quantum chip technology. To achieve this goal, it is crucial to prove that PDMR readout is compatible with the coherent spin control. Here we present PDMR MW pulse protocols that filter background currents related to ionization of $N_s^0$ defects and achieve a high contrast and S/N ratio. We demonstrate Rabi and Ramsey protocols on shallow nitrogen-implanted electronic grade diamond and the coherent readout of ∼ 5 NV spins, as a first step towards the fabrication of scalable photoelectric quantum devices.

**Main text:**

Hybrid quantum detection approaches are nowadays considered as the most promising towards the realisation of scalable systems for quantum computation (1). Recently we have introduced the photoelectrically detected magnetic resonance (PDMR) technique (2) for reading out the electron spin state of negatively charged nitrogen-vacancy (NV) centre in diamond. This technique has a potential to be used in scalable miniaturized electric device technology (3), going thus beyond the abilities offered by optically detected magnetic resonance (ODMR) (4). Further on, the PDMR technique provides a vast enhancement of over 2.5 orders of magnitude in detection rates compared to ODMR measured in similar experimental conditions (5). Although we have demonstrated the principle of this method, it was only used so far for CW readout of the |0> and |±1> spin states of NV ensembles in lab-grown CVD or HPHT diamonds (2). In order to use the advantages of the PDMR technique in quantum sensing and computational schemes, it is necessary to demonstrate that the photoelectric readout is compatible with the quantum control of individual NV electron spins and with pulsed magnetic resonance experiments enabling preparation of arbitrary quantum states of NV centres. In this work, we demonstrate such quantum ability to perform coherent operations with PDMR on spin ensembles of just a few (∼ 5) NV by designing specific readout sequences. We demonstrate these protocols by performing Rabi and Ramsey measurements on shallow NV ensembles implanted in electronic grade diamond, up to high frequencies of ~ 20 MHz. The proposed method leads also to a significant suppression of background photocurrents, which is one of the main limiting factors for using PDMR for readout of single NV sites (2; 6). The sensitivity of this method allows us to perform coherent operations with photoelectric readout on small spin ensembles with considerably increased signal-to-noise (S/N) ratio compared to CW PDMR measurements (2; 5). By this way we could experimentally verify the S/N ratio that were previously only theoretically projected (7). We discuss the detection noise and show that a shot noise < 0.18 fA·Hz$^{-1/2}$ is achieved for the used photo-ionization rate.

Figure 1a depicts the hybrid photoelectric quantum chip used in our measurements. Type-IIa single crystal electronic grade diamond was implanted with 8 keV $^{14}N^{4+}$ ions and annealed at 900°C to create ensembles of shallow NV centres (depth: 12 ± 4 nm). The implantation fluence was varied from $1 \times 10^{10}$ to $1 \times 10^{14}$ $N^{14+} \cdot cm^{-2}$ to create in total five regions with different NV densities. By this way a density of NV centres from ~ 15 µm$^{-2}$ to 10$^4$ µm$^{-2}$ was obtained. The sample was equipped with coplanar Ti-Au electrodes with 50 µm gap and mounted on a circuit board to enable microwave (MW) excitation (8) and photocurrent readout. MW was provided using two printed antennas connected in series and terminated by 50 Ω (see SI for details). Green laser illumination is used to induce the two-photon ionization of NV centres from the |0> and |±1> $^3A_2$ NV ground state (GS) spin manifolds to the conduction band (CB) (Figure 1b) (9; 10). A DC electric field is applied in between electrodes and the photocurrent is readout by lock-in amplification after sensitive pre-amplification (see SI).

The magnetic resonance (MR) contrast in the detected photocurrent is obtained under resonant microwave excitation (2.87 GHz) inducing transitions from the |0> to the |±1> spin sublevels of NV ground state (11). Electrons photo-excited from the |±1> GS manifold have a non-zero probability to undergo shelving transitions from NV excited state to the metastable $^1E$ state (12; 13; 14). During the time in which the |±1> spin electron is stored in the metastable singlet state (~ 220 ns lifetime) (2; 15; 16), it is protected from photo-ionization. This leads to a reduction of the overall photocurrent and thus to a minimum in the photocurrent signal at the resonant MW frequency. However, the green laser excitation used for PDMR photo-ionizes other defects such as substitutional nitrogen ($N_S^0$), providing a constant photocurrent background (5). This background reduces the detected MR contrast and limits the application of PDMR for single NV detection and. So far, the photocurrent signal from NV centres was readout either by lock-in amplification referenced to the laser pulse frequency (2; 5) or by detecting the DC photocurrent after implementing a laser readout pulse (7). Both proposed methods suffer from the mentioned background photocurrent limiting the S/N ratio. Here we perform PDMR readout using pulse sequences and referencing the lock-in signal to the MW pulses carrier frequency. By this way, we detect only the proportion of the photocurrent affected by the MW field, i.e. variations in the photocurrent related to |0> ↔ |±1> transitions. The background signal from other defects than NV is filtered out when scanning the MW frequency around NV resonant line. The measurement contrast between off (|0>) and on spin resonance (|±1>) photocurrent is therefore practically 100%. To quantify the benefit of MW-triggered PDMR, we compared the S/N ratio for CW PDMR measurements performed with laser- and MW-triggering of the lock-in amplifier at the same position (highly implanted region of the sample, corresponding to ~ 1000 NVs in focus of the objective) and in similar experimental conditions. The photocurrent signal S is determined from a fit of PDMR spectra. The noise N is calculated as the standard deviation of the blank photocurrent signal outside of the magnetic resonance. When comparing the calculated values from MW and laser triggered PDMR we get a 7.4 times improvement in S/N for the MW-triggering method.

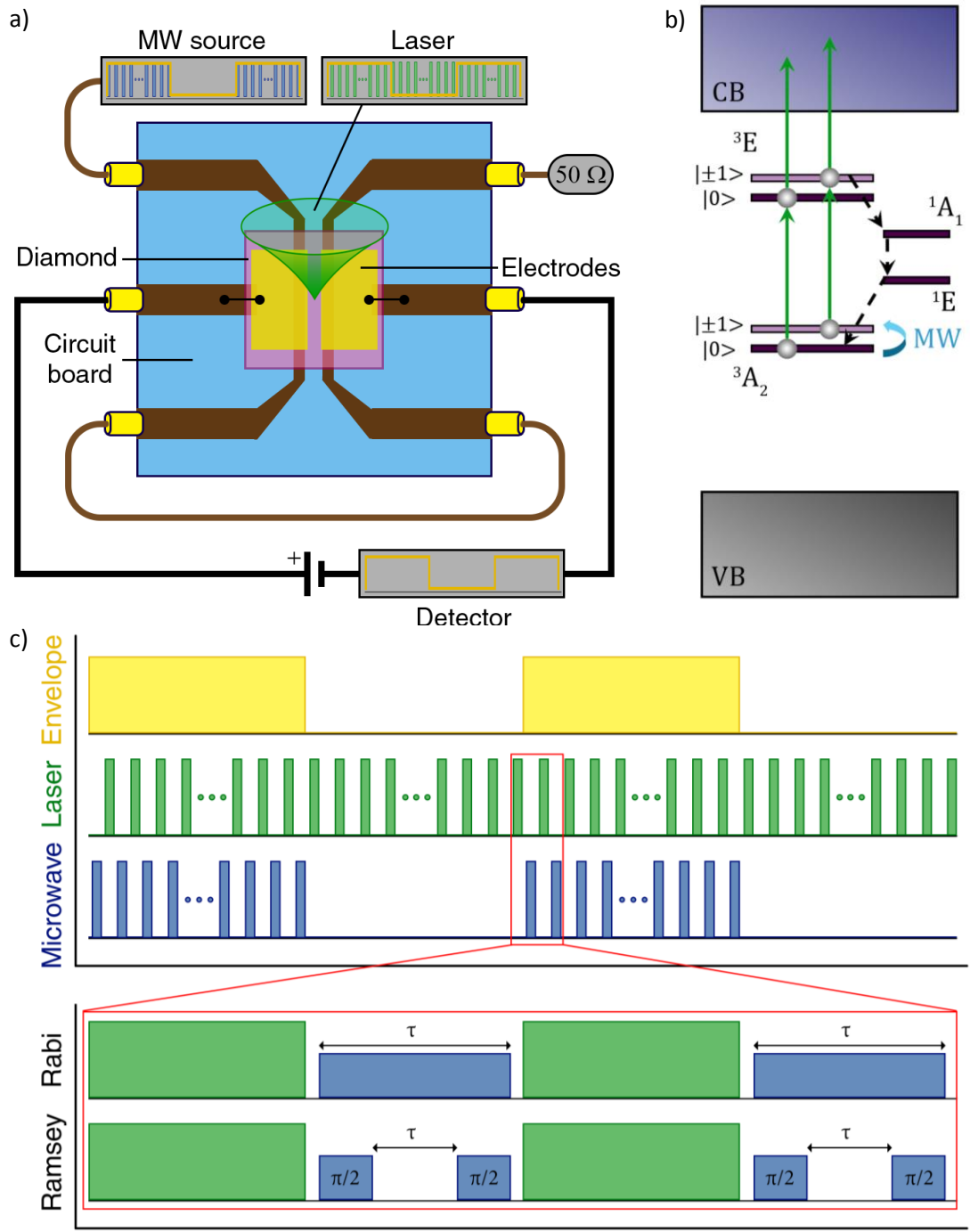

*Figure 1: (a) The schematics of the setup used for pulsed PDMR showing a type-IIa nitrogen implanted diamond mounted on the circuit board to enable microwave excitation and current readout. (b) A simplified electronic energy level scheme of the NV centre showing two-photon ionization of electron from GS to conduction band (CB). Spin-selective decay through metastable $^1E$ state enables the PDMR*

*(MW: microwave, VB: valence band). (c) Sequences used for pulsed PDMR that encode high frequency pulse sequences (Rabi, Ramsey) into a low frequency envelope used as a reference for lock-in detection.*

To achieve coherent manipulation of spins with photoelectric readout, which is crucial to enable single NV spin photoelectric detection, we have designed pulse sequences. In our pulsed protocols, we encode single pulse sequences into a low-frequency envelope, as schematically shown in Figure 1c. For this, a continuous series of consecutives laser pulses is used, in which each pulse serves for the spin initialization and spin state readout. Bursts of spin manipulation MW pulses, time-shifted with respect to laser pulses, are added to encode any arbitrary sequence. A MW trigger pulse marks the start of the pulse burst during which the low frequency modulated photocurrent signal is measured by the lock-in amplifier. In the off period of the duty cycle, a train of spin polarisation laser pulses is used to keep the occupation on the 0> state. To obtain the best S/N ratio, we optimize the laser pulse length (∼ 800 ns) and the delay between laser and MW π-pulses (200 to 300 ns in the case of the measured sample). We explored envelope frequencies from 30 Hz up to 1 kHz, tuned to obtain the highest S/N ratio. In principle, this method enables encoding any pulse sequence into the low-frequency envelope. Here we demonstrate this technique on Rabi and Ramsey measurements, which are key measurements for magnetometry and other quantum measurement applications (17).

Rabi oscillation measurements on ensemble of NV spins (see Figure 2) were performed first in the region of the sample with an approximate number of 1000 NV centres in the focus of the objective (region R1, see SI). To determine the duration of π-pulses (MW pulse that flips the NV spin orientation from |0> to |±1> state and vice versa) for specific MW powers, we split the |±1> spin energy levels of NV centres by applying a static magnetic field perpendicular to the <100> diamond plane and we induce the oscillations of population between the |0> and |-1> spin sublevels. Laser pulses serve for both initialization of the spin state in the |0> spin sublevel and for readout of the NV centre spin state. We then vary the duration of resonant MW pulses at constant power (see Figure 1c) to detect the Rabi oscillations. The obtained data is fitted using the formula (18)

$$F(\tau_{RAB}) = A[1 - B \cdot \cos(2\pi\nu_{RAB}\tau_{RAB} + \phi)]e^{-\frac{\tau_{RAB}}{T_2}} + C,$$

where $\nu_{RAB}$ is the Rabi frequency, $\phi$ is the phase shift, $\tau_{RAB}$ is the MW pulse duration, $T_2$ is the approximate transverse relaxation time and *A, B, C* are the fitting parameters. To obtain high S/N ratio we average over $2\times10^6$ Rabi sequences. As expected, the Rabi frequencies calculated from the fit of these measurements depend linearly on the square root of the MW power (Figure 2 inset). The presented measurement proves the ability of coherent spin readout using MW triggered PDMR envelope pulse scheme. The measured Rabi oscillations for MW power of 3.2 W are plotted in Figure 2, yielding $T_2 \sim$ 490 ns for the given sample on the area with the highest $^{14}N^{4+}$ ions implantation fluency.

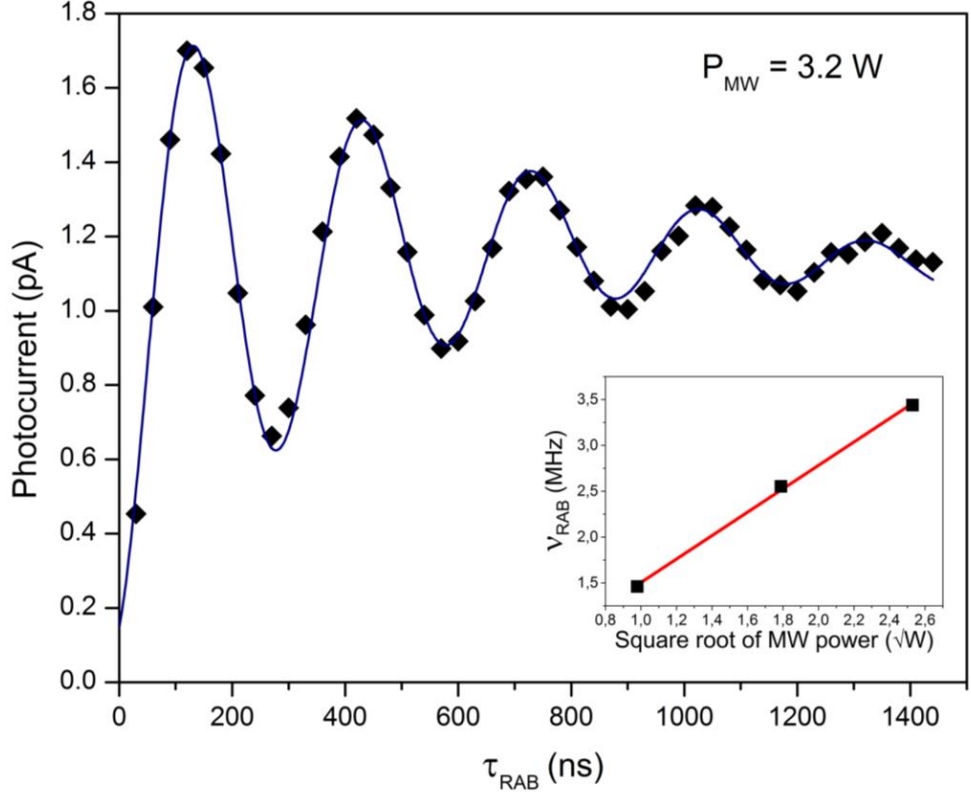

*Figure 2:* *Photoelectric detection of Rabi oscillations using envelope technique with microwave triggering. The black scatter points are the experimental data for MW power of 3.2 W and the blue line is fitted. Inset shows the dependence of the frequency of Rabi oscillations with respect to the square root of the applied microwave power (red line is the result of linear fit).*

To show the applicability of the designed PDMR quantum protocols, we demonstrate functional Ramsey fringes pulse sequences (Figure 1c), employed to characterize the NV centre free spin dynamics. The measurement is performed in the same conditions as for the Rabi experiment. We use laser pulses to initialize and readout the NV spin state. In between laser pulses, we implement a Ramsey sequence. In this sequence, a first π/2 MW pulse (duration of ∼ 30 ns at 16 W, determined from Rabi measurements) is applied to prepare the NV electron spin in a state of superposition of the |0> and |-1> spin sublevels. After a duration $\tau_{RAM}$ during which the spin can freely precess and dephase, a second π/2 MW pulse is used to project the spin back into the |0> and |-1> basis. By varying the free precession time $\tau_{RAM}$ we observe a decay curve consisting of the superposition of three cosine functions with different frequencies $\delta_i$. Each frequency corresponds to the detuning of the applied MW frequency from one of the hyperfine $^{14}$N transition such that $\delta_1$ = $\nu_{RAM}$ - 2.2 MHz, $\delta_2$ = $\nu_{RAM}$ and $\delta_3$ = $\nu_{RAM}$ + 2.2 MHz (19; 20), where $\nu_{RAM}$ is the Ramsey frequency. We have fitted the data considering a single exponential decay (21) with a function

$$F(\tau_{RAM}) = A + B \cdot \tau_{RAM} + C \cdot exp[-\tau_{RAM}/T_2^*] \sum_{i=1}^{3} \cos(2 \cdot \pi \cdot \delta_i \cdot \tau_{RAM} + \phi_i),$$

where $\delta_i$ are the oscillation frequencies, $\phi_i$ are the phase shifts (20), $\tau_{RAM}$ is the free precession time, $T_2^*$ is the spin dephasing time and *A, B, C* are the fitting parameters. The measurement was performed for different detunings $\delta_2$ (4, 8, 12, 16, 20 MHz) from the central resonant frequency. The results are shown in Figure 3. As expected we observe a linear dependency of the Ramsey frequency (obtained from the fit) to the different detuning values (Figure 3 inset). The average dephasing time $T_2^*$ for all measurements is ∼ 41 ns using single exponential fitting or ∼ 66 ns using Gaussian fitting of the data (22). This short value reflects decoherent interactions of NV ensembles on the sample area over-implanted with $^{14}N^{4+}$ ions (region R1, see SI). The main source of dephasing is the bath of paramagnetic substitutional nitrogen impurities ($N_S^0$ centres) and the high concentration of $^{14}N$ nuclear spins in diamond samples with high nitrogen concentration (15; 23). Despite the very fast decay, our readout technique enables to readout Ramsey fringes at high frequency of 20 MHz with a high contrast, confirming the ability of MW-referenced PDMR to measure samples with very short $T_2^*$ times with high S/N ratio. By the proposed technique, Hahn-echo or more complex sequences such as CPMG can be encoded into the envelope carrier. The base value of the MW triggered photocurrent is inversely proportional to the detuning frequency, reflecting the occupation of spin states integrated over the statistical weight of the |0> and |-1> manifold.

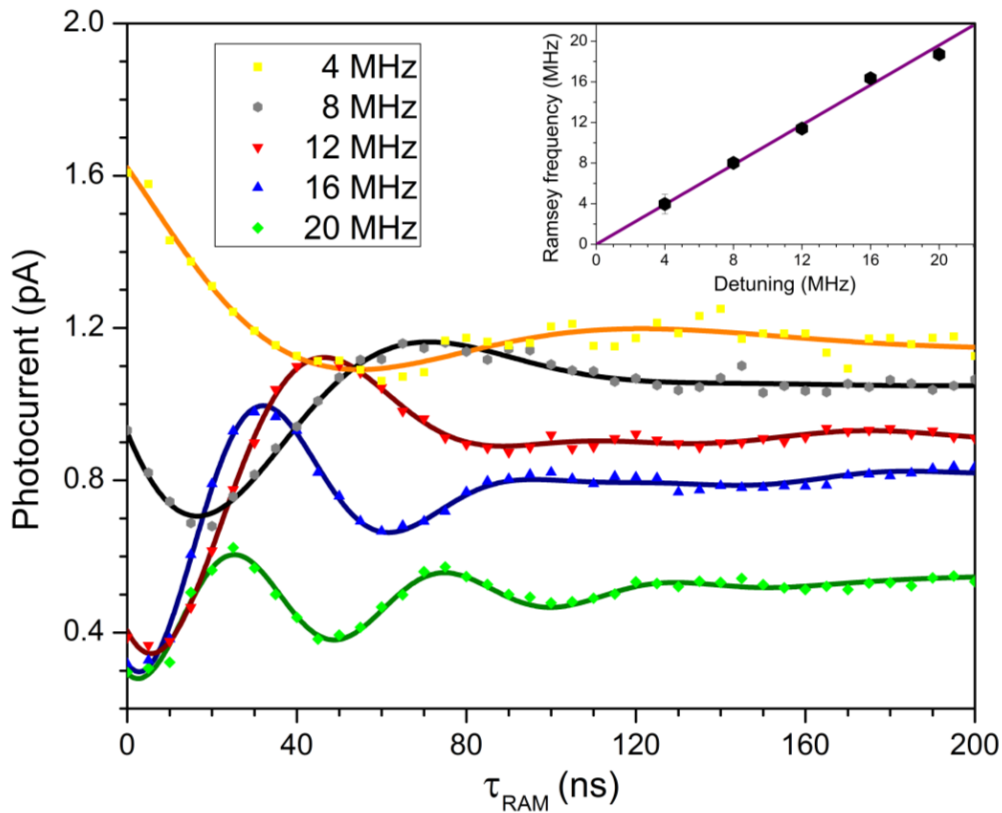

**Figure 3:** *Photoelectric detection of Ramsey fringes for different detuning from MW excitation frequency from the NV resonant frequency. The scatter points are the experimental data, fitted considering an exponential decay (continuous lines). Inset shows the linear dependence of Ramsey frequency to the MW detuning (purple line is the result of linear fit).*

An important milestone for the PDMR technique is to reach coherent manipulation of single NV spins with photoelectric readout. While our recent studies (2; 5) and other published work (7) demonstrated PDMR on NV ensembles, the photoelectric readout of single spins was not achieved yet. To study systematically the downscaling of the number of NV centres in focus of the objective and corresponding PDMR signal, we used the N-implanted electronic grade diamond containing five regions of descending NV densities, corresponding to ensembles from 1000 to ∼ 5 NV centres in focus of the objective. The number of NV centres in the focus was calibrated from photoluminescence counts of a single NV obtained in same conditions.

First we measure the total photocurrent signal on NVs ensembles and systematically downscale the number of NV centres in the focus. For this we use lock-in amplification referenced to a pulsed laser, without the MW field applied (see Figure 4 inset). This data could be fitted using a sublinear function. We speculate that this is due to a reduction of recombination lifetime in highly implanted region. From the fit we predict the total photocurrent resulting from the ionization of a single NV to be ∼ 1.0 pA for an electric field of $2 \times 10^4$ V·cm$^{-1}$, similarly to the prediction presented in (2; 7). In this case the measured photocurrent is a sum of NV-related photocurrent and of the background signal originating mainly from single substitutional nitrogen $N_s^0$ ionization. We estimate that at these power conditions ∼ 75% of the signal comes from two-photon ionization of the NV centres (5). Here we could experimentally determine for the first time the photocurrent expected for single NV centers at the given electric field conditions by measurements on a small number of NV centers (∼ 5), downscaled by at least ∼ $10^4$ times compared to previous measurements (7).

To read precisely the photocurrent related to NV centre in pulsed schemes without any background and to achieve the coherent spin manipulation and readout on small NV ensembles, we used pulse PDMR protocols with MW triggering. By measuring the Rabi oscillations, we determine the length of π-pulses (70 ns at 4.9 W of MW power) which are used to induce resonant |0> ↔ |±1> transitions (with no external magnetic field applied). Figure 4 shows PDMR spectra obtained for decreasing numbers of NV centres in the focus of the objective. Data were fitted using a two-peak Gaussian function (24), averaging over the central and hyperfine MR lines. The estimated value for single NV photocurrent at resonance frequency averaged from all four measurements is ∼ 2 fA. This value is ∼ 50 times lower than the photocurrent expected for a single-NV in the used conditions based on CW light-triggered measurements (see above) and considering a PDMR contrast of ∼ 10 % (5). This is attributed to several factors such as the use of duty cycle (resulting in a reduced effective MW and laser power) and the loss of the contrast during the laser readout pulse (since the laser pulse is used for both readout and initialization of the NVs). However as shown in Figure 4, the pulsed photocurrent related to the magnetic resonance of ∼ 5 NV centres is detected with very high contrast.

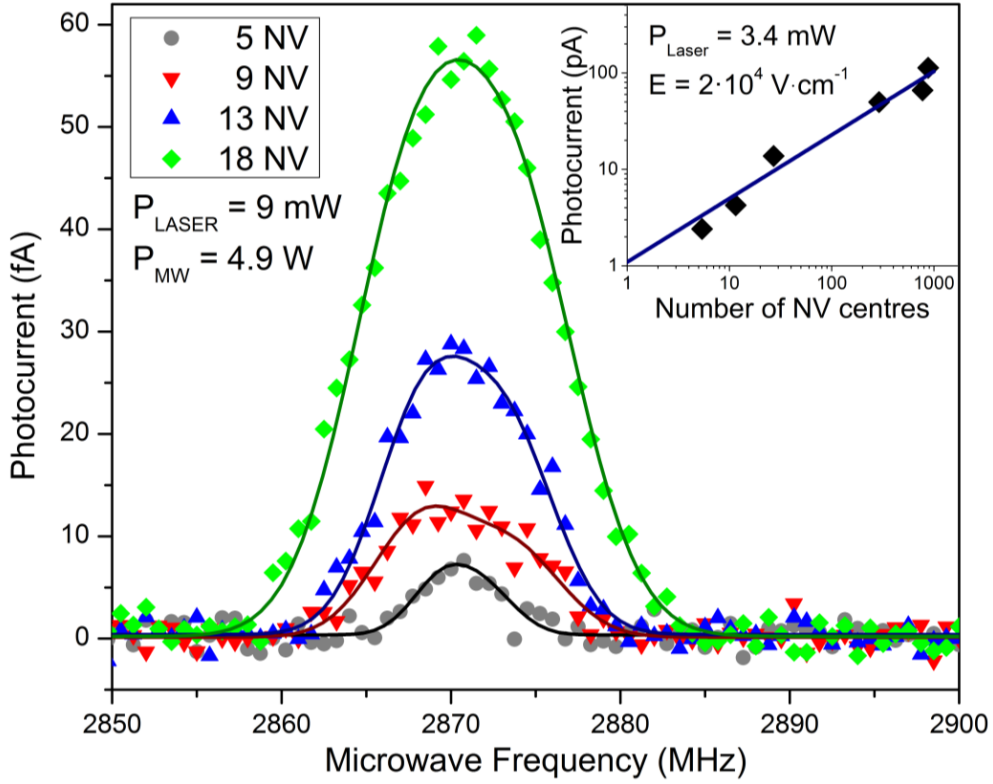

*Figure 4: Data measured on small spin ensembles using a sequence consisting of short laser pulse followed by a MW π-pulse. The experimental data are fitted with two Gaussian functions. For comparison the inset shows the total photocurrent from different numbers of NV centres measured with only laser excitation (blue line is the result of fit).*

To explore the limitations of the pulsed PDMR method we concentrated on the comparison of theoretical and experimental values of noise level. The main components of the total noise $N_{rms}$ come from the shot noise generated by the photocurrent $I_{SN}(\text{rms})$ (25; 26), input noise of the pre-amplifier $I_{Preamp}(\text{rms})$ and input noise of the lock-in amplifier $I_{Lock-in}(\text{rms})$, expressed as

$$N_{rms} = (\delta I_{SN}(rms) + \delta I_{Preamp}(rms) + \delta I_{Lock-in}(rms)) \sqrt{\Delta f}$$

where Δf is the bandwidth of detection electronics. Thanks to the lock-in amplifier detection, one is able to reduce the noise significantly, since in this case Δf corresponds to the equivalent noise bandwidth (ENBW) determined by the lock-in amplifier time constant. The ENBW can be then very narrow, thus increasing the S/N by rejecting the white noise (25). The total noise is then calculated as

$$N_{rms} = (\sqrt{2 \cdot e \cdot I} + \delta I_{Preamp}(rms) + \delta I_{Lock-in}(rms)) \cdot \sqrt{\text{ENBW}}$$

where *e* is the elementary charge and *I* is the measured photocurrent. Considering the specifications for our preamplifier and lock-in amplifier, we calculate for the CW technique $N_{rms}/\sqrt{\text{ENBW}}$ of ~ 60.6 fA·Hz$^{-1/2}$, with a dominant noise originating from the preamplifier (~ 60 fA·Hz$^{-1/2}$ at given frequency), compared to measured value of ~ 83.7 fA·Hz$^{-1/2}$ for 1000 NVs in the focus. We speculate that this difference (23.1 fA·Hz$^{-1/2}$) between the experimental and predicted noise is due to photocurrent noise correlated to MW cross-talk occurring on the chip level. However this type of cross-talk noise should be significantly reduced for the pulsed measurements where the laser readout pulse is decoupled from MW pulse. To confirm this we have determined the correlated noise for the pulsed experiment in Figure 4 using the pulsed protocols. In this configuration the correlated noise is reduced to 4.4 fA·Hz$^{-1/2}$. The correlated noise compared to the optically triggered CW PDMR is then reduced by a factor of 48. The shot noise for the MW triggered PDMR is ~ 0.179 fA·Hz$^{-1/2}$ calculated for a single NV spin photocurrent (considering a differential current between spin |0> and |±1> states of 100 fA).

To conclude, in this letter we demonstrate coherent readout techniques for small ensembles of individual NV centres. Several drawbacks had to be overcome to enhance the signal-to-noise ratio. This has been achieved by designing MW referenced pulse sequences, enabling to eliminate the $N_s^0$-related background photocurrents, which was identified as a main obstacle for using PDMR technique for readout of single NV spins. Further on, the significant enhancement of S/N ratio compared to optically referenced technique made possible the development of coherent spin manipulation pulse PDMR protocols encoding high frequency MW and laser pulse sequences into a low frequency envelope, reducing substantially the correlated noise. Pulsed spin manipulation Rabi and Ramsey protocols were demonstrated on ensembles of shallow NV centres engineered in electronic grade diamond, which is the material used for relevant quantum applications. The developed protocols allowed to control ~ 5 NV centres in the focus of the objective. A further enhancement in S/N ratio can be achieved by designing nanoscopic electrodes for operating efficiently single NVs with high photoelectron gain (2; 5). The data presented here demonstrate perspectives for progressing towards compact and scalable single spin device architecture on an electrical quantum chip.

**Acknowledgements:**
Support from EU (FP7 project DIADEMS, grant No. 611143), Grant Agency of the Czech Republic (Project No. 16-16336S) and from bilateral project BOF15BL08 between University of Hasselt and Czech Technical University in Prague is acknowledged. We would like to thank Professor Dmitry Budker from Johannes Gutenberg University, Mainz and from University of California, Berkeley for valuable discussions.

**Supplemental Material:**

Sample preparation:
For the experiment we used a chemical vapour deposited (CVD) electronic grade type-IIa diamond plate purchased from Element Six. The sample was implanted with 8 keV $^{14}$N$^{4+}$ ions at the Electron Cyclotron Resonance Ion Source Sophie (Vienna University of Technology). The implantation fluency varied for different regions of the sample to create in total 5 different regions denoted R1, R2, R3, R4 and R5 with fluency of 1.00·10$^{14}$, 1.00·10$^{12}$, 0.96·10$^{10}$, 0.99·10$^{11}$ and 1.01·10$^{13}$ N$^{4+}$/cm$^2$ respectively. After implantation, the sample was annealed at 900°C in argon atmosphere to create ensembles of shallow NV centres (depth: 12 ± 4 nm). When illuminated with green laser (10 mW), the average photon count is 180.7 Mcts/s for the region R1 with the highest implantation density in comparison to 185.3 Mcts/s for the region R5. This observation suggests the over-implantation of the R1 region and creation of high content of N-related defects other than NV centres.

Laser and MW excitation:

For green light excitation we use a GEM Nd:YAG laser (532 nm) set to 9 mW, focused on the sample with a NA = 0.95, 40x air objective, creating a light spot of about 680 nm diameter. For MW excitation we use two printed parallel antennas (optimized for NV resonance frequency) connected in series and terminated by 50 Ω (8). The antennas are oriented alongside the readout electrodes described below (see Figure 1a in main text). Microwaves are provided by RF-signal generator (TEG4000-1 from Telemakus) with an output power of 1 mW. The final microwave power is set to the powers ranging up to 16 W using an attenuator (TEA4000-3 from Telemakus) and broadband amplifier with a gain of 45 dB (ZHL-16W-43+ from Mini-circuits).

Photocurrent detection:

Coplanar readout electrodes with a gap of 50 μm were prepared on the diamond crystal by photolithography and sputtering of 10 nm of titanium topped by 100 nm of gold. Electrodes are connected to the circuit board using wire bonding and a DC voltage (90 to 100 V corresponding to $1.8 \times 10^4$ to $2 \times 10^4$ V·cm$^{-1}$) is applied to drive the photo-generated charge carriers toward electrodes. Photocurrent was pre-amplified using Standford Research SR570 pre-amplifier and subsequently readout by SR850 lock-in amplifier.

CW and pulse measurements:

For CW PDMR operation, the laser was set to CW mode and MW were pulsed with triggering frequency of 37 Hz (supplied by National Instrument PCI card). The lock-in amplifier was referenced to the same frequency. In case of pulsed schemes, an envelope signal of 211 Hz (611 Hz for measurements on small NV ensembles) was used as the envelope wave, referenced to the lock-in amplification. The low frequency driving envelope was provided from the computer clock. Two arbitrary waveform generators (AWG) were used to control respectively the acousto-optic modulator for laser pulsing and the MW switch for MW pulsing for implementation of the pulse protocols encoded to the low frequency envelope. To fix the phase shift between the laser and MW pulses, corresponding delays were set on individual AWGs. The duration of laser pulses was set after experimental optimization to typically 800 ns (short enough to provide sufficient contrast, long enough to initialize the ensemble of NVs). The delay between laser and MW pulses was set to 300 ns for Rabi and 200 ns for Ramsey experiment. To reference the lock-in amplifier to the MW signal, the total number of laser pulses was set to fill up the whole period of the envelope triggering signal, whereas the number of pulses for MW was set to fill up only the half of the envelope triggering signal period (see Figure 1c in main text).